\documentclass[sigconf]{acmart}
\AtBeginDocument{%
  \providecommand\BibTeX{{%
    \normalfont B\kern-0.5em{\scshape i\kern-0.25em b}\kern-0.8em\TeX}}}

\copyrightyear{2024}
\acmYear{2024}
\setcopyright{acmlicensed}\acmConference[ICSE-SEET '24]{46th International Conference on Software Engineering: Software Engineering Education and Training}{April 14--20, 2024}{Lisbon, Portugal}
\acmBooktitle{46th International Conference on Software Engineering: Software Engineering Education and Training (ICSE-SEET '24), April 14--20, 2024, Lisbon, Portugal}
\acmDOI{10.1145/3639474.3640081}
\acmISBN{979-8-4007-0498-7/24/04}

\usepackage{multirow}
\usepackage{acronym}
\usepackage{amsthm}

\usepackage{tabularx}
 \newenvironment{conditions*}
  {\par\vspace{\abovedisplayskip}\noindent
   \tabularx{\columnwidth}{>{$}l<{$} @{${}={}$} >{\raggedright\arraybackslash}X}}
  {\endtabularx\par\vspace{\belowdisplayskip}}

\newacro {OSS}[OSS]{Open Source Software}
\newacro {PR}[PR]{Pull Request}
\newacro {RAT}[RAT]{Readiness Assurance Test}
\newacro {TBL}[TBL]{Team-based Learning}
\newacro {TDD}[TDD]{Test-Driven Development}
\begin{document}

\title[Bridging Theory to Practice in Software Testing Teaching]{Bridging Theory to Practice in Software Testing Teaching through \ac{TBL} and \ac{OSS} Contribution}
 \acresetall

\author{Elaine Venson}
\orcid{0000-0002-7607-5936}
\affiliation{%
  \institution{University of Brasilia}
  \city{Brasilia}
  \country{Brazil}
}
\email{elainevenson@unb.br}

\author{Reem Alfayez}
\orcid{0000-0001-6782-247X}
\affiliation{%
  \institution{College of Computer and Information Sciences \\King Saud University}
  \city{Riyadh}
  \country{Saudi Arabia}}
\email{reealfayez@ksu.edu.sa}

\renewcommand{\shortauthors}{Venson and Alfayez}

\begin{abstract}
Curricula recommendation for undergraduate Software Engineering courses underscore the importance of transcending from traditional lecture format to  actively involving students in time-limited, iterative development practices. This paper presents a teaching approach for a software testing course that integrates theory and practical experience through the utilization of both \ac{TBL} and 
active contributions to \ac{OSS} projects. The paper reports on our experience implementing the pedagogical approach over four consecutive semesters of a Software Testing course within an undergraduate Software Engineering program. The experience encompassed both online and in-person classes, involving a substantial cohort of over 300 students spanning four semesters.
Students' perceptions regarding the course are analyzed and compared with previous, related studies. Our results are positively aligned  with the existing literature of software engineering teaching, confirming the effectiveness of combining \ac{TBL} with \ac{OSS} contributions. Additionally, our survey has shed light on the challenges that students encounter during their first contribution to \ac{OSS} projects, highlighting the need for targeted solutions.
Overall, the experience demonstrates that the proposed pedagogical structure can effectively facilitate the transition from 
 theoretical knowledge to real-world practice in the domain of Software Testing.

\end{abstract}
 \acresetall




\maketitle

\section{Introduction}


The Association for Computing Machinery (ACM)'s curriculum guideline states that software engineering education needs to move beyond the traditional lecture format and consider the incorporation of a variety of teaching and learning approaches \cite{joint_task_force_on_computing_curricula_ieee_computer_society_association_for_computing_machinery_curriculum_2014}. Yet, software engineering instructors encounter significant challenges in software engineering courses when it comes to effectively engaging students and facilitating the transition from theoretical concepts to practical application \cite{ouhbi_software_2020}.  

\ac{TBL} is an instructional strategy and active learning approach designed to enhance student engagement and promote collaborative learning. It is often used in higher education settings, particularly in courses where problem-solving, critical thinking, and teamwork are important learning outcomes. In TBL, students work in teams to address complex, real-world problems and case studies \cite{michaelsen_essential_2008}.

\ac{TBL} has demonstrated its effectiveness in engaging a new generation of students who increasingly resist traditional expository lecture-style classes, yielding positive results \cite{matalonga_deploying_2017, van_helden_collaborative_2023}. The utilization of \ac{TBL} not only demonstrated a positive impact on students' performance and grades, as evidenced by \cite{joshi_evaluating_2020, nawabi_team-based_2021}, but it has also revealed that courses employing this approach are well evaluated by students \cite{matalonga_deploying_2017, nawabi_team-based_2021}.


A course utilizing the \ac{TBL} approach will structure its content into modules, typically five to seven. Each module consists of the same three primary stages.
The first two stages are: "Preparation" and "Readiness Assurance", and they are focused on acquiring theoretical knowledge about the topic being studied. The third stage is "Application of Course Concepts" , which involves in-class activities and assignments designed for students to apply and practice the course content \cite{michaelsen_essential_2008}.

Proposing assignments that are aligned with
real-world problems is an important aspect in Software Engineering courses \cite{joint_task_force_on_computing_curricula_ieee_computer_society_association_for_computing_machinery_curriculum_2014}. Recently, computer science educators have been taking advantage of public software development repositories such as GitHub to provide such experiences to students \cite{smith_selecting_2014, deng_teaching_2020, he_open_2023}. For example, the study of Deng et al. \cite{deng_teaching_2020} proposes five assignments for a Software Testing course that utilize Free and \ac{OSS} projects to provide students with real-world hands-on software testing experience.

This paper reports on our experience applying the \ac{TBL} approach and \ac{OSS} contributions  in a Software Testing undergraduate course. Extending \ac{TBL}application activities, students were requested to participate in 
\ac{OSS} through software testing assignments based on Deng et al.'s work \cite{deng_teaching_2020}. We divided the Software Testing course content into five modules that each includes an \ac{OSS} assignment. In particular, 
each module begins with students acquiring theoretical knowledge about a specific software testing topic. Subsequently, they transition to practical application through the engagement in classroom exercises that are focused on the topic learned, aiming to reinforce the studies concepts. The module concludes with students contributing to an \ac{OSS} project. We evaluated the effectiveness of the approach by gauging student perceptions of each \ac{TBL} activity in addition to the overall approach across four semesters, spanning eight sections and encompassing over 300 students, from 2021 to 2023.

\section{Background and Related Work}
\subsection{Team-Based Learning (TBL)}
The primary objective of \ac{TBL} is to shift the classroom experience from knowledge acquisition to knowledge application. Most of the learning takes place among students within their groups, while the instructor is consistently present and available to provide guidance on the material and content being studied \cite{michaelsen_essential_2008}.

The \ac{TBL} approach is structured around instructional units called ``modules''. These modules follow a three-step cycle, encompassing preparation, in-class readiness assessment testing, and application-focused exercises. 
The \textit{Preparation} phase marks the beginning of each module, during which students study assigned materials, which could include text, audio, or videos
The subsequent phase, known as the Readiness Assurance, comprises an individual multiple-choice test, followed by a team test. In the team test, the members collaborate to answer the same set of questions, engaging in discussions to reach a consensus. Immediately following the in-class tests, teams have the opportunity to submit appeals if they disagree with the answers to specific test questions. These appeals serve as an additional learning opportunity, as teams must construct requests based on their study materials. The final phase of each module, the Application of Course Concepts, entails teams collaborating to solve problems by applying the knowledge they have acquired throughout the module \cite{michaelsen_essential_2008}.

According to Michaelsen and Sweet \cite{michaelsen_essential_2008}, four essential elements are required for the successful implementation of \ac{TBL}: well-formed and managed groups, individual and collective accountability, consistent feedback, and assignments designed to enhance learning and team development.

Studies indicate that students in disciplines that employ \ac{TBL} learn as much or even more content and concepts compared to those in lecture-based courses \cite{matalonga_deploying_2017, van_helden_collaborative_2023}. Furthermore, this teaching and learning method also fosters the development of teamwork skills and the application of course content in real-world situations \cite{joshi_evaluating_2020}.

\subsection{\ac{TBL}Applications in Computer Science Courses}

The \ac{TBL} model has been applied in Computer Science and Software Engineering, spanning undergraduate and graduate level courses.

The application of \ac{TBL} to a Software Engineering course in the University of ORT
in Uruguay received a general positive evaluation by students and instructors \cite{matalonga_deploying_2017}. Considering that designing the application exercises is a key element for the successful deployment of the 
methodology, their study applies a combination of multiple choice questions, serious games, and open-ended exercises to motivate students to participate.

At Boise State University, \ac{TBL} was implemented in a Systems Programming course, leading to statistically significant enhancements in student performance in particular assignments and fostering a stronger sense of community among students.This was reflected in an 88\% positive response rate in an end-of-semester survey, ultimately resulting in higher overall class grades when compared to a traditional lecture-based approach, as detailed in \cite{joshi_evaluating_2020}.

An experience report on the implementation of the Flipped Classroom (FC) and \ac{TBL} in an introductory programming course at Reykjavik University yielded mixed student feedback. Notably, 47\% of the students expressed satisfaction with the course organization, and 60\% had a favorable opinion of \ac{TBL}. However, 33\% of the students conveyed dissatisfaction, particularly among female students, and roughly half of the students expressed a preference for traditional lectures. Surprisingly, a significant portion of the students (44\%) did not read the textbook, while a larger majority (74\%) did watch the provided videos \cite{loftsson_using_2019}.

 Awatramani and River implemented \ac{TBL} in a graduate computer engineering course focused on system-level design of embedded systems \cite{awatramani_team-based_2015}. The authors evaluated the course content using the Florida Taxonomy of Cognitive Behavior, aiming to gauge the effectiveness of their \ac{TBL} implementation in fostering deeper learning. Their findings indicated that they surpassed the Bloom's application level in some exercises; however, the overall cognitive assessment score fell short of their expectations. The authors are optimistic that with more experience in using \ac{TBL} and by placing greater emphasis on higher-order skills in course design, along with the incorporation of feedback and reflection practices into the assessment process, they can enhance the scores.

Presler-Marshall et. al. \cite{presler-marshall_improving_2023} reported on the challenges 
and improvements in grading practices within \ac{TBL}in the context of an introductory software engineering course. The study introduces an algorithm and a tool named AutoVCS to automate the assessment of individual student contributions in GitHub repositories. AutoVCS resulted in improved grading consistency and higher-quality feedback for teaching assistants. This led to a strong preference among teaching assistants for automated summaries. However, the tool did not significantly expedite the grading process.

\subsection{Open-Source Software (OSS) in Software Engineering Education}

\ac{OSS} serves as a valuable resource in Software Engineering education, providing real-world examples of code that closely resemble the challenges encountered by software engineers in practical settings. For example, utilizing \ac{OSS}-based Software Engineering projects exposes students to the task of deciphering poorly documented or badly written code, simulating industrial challenges within an academic setting \cite{smith_selecting_2014}.

Several mapping studies aimed to understand the varying aspect of utilizing \ac{OSS} in software engineering education \cite{brito_floss_2018}.

Spinellis \cite{spinellis_why_2021} advocates for the active involvement of computing students in contributing to \ac{OSS} projects as an integral part of their academic curriculum. With more than 15 years of experience in teaching a software engineering course that mandates \ac{OSS} contributions, the author underscores how this practice has transformed into a contemporary extension of skill acquisition beyond coding. The paper provides insights into the valuable learning outcomes that students gain from such engagement, elucidates the integration of \ac{OSS} contribution exercises within the course, and concludes by outlining the best practices that have contributed to the success of this practice.

To gain an understanding of educators' perspectives on incorporating \ac{OSS} projects into their education practices, Pinto et al. \cite{pinto_training_2017}
conducted interviews with seven Software Engineering professors who have integrated \ac{OSS} into their teaching. The study revealed that there are diverse approaches to incorporating \ac{OSS} projects in software engineering courses, including project selection, assessment methods, and learning objectives. The study highlights benefits resulting from this pedagogical shift, such as the improvement of students' social and technical skills and the valuable enhancement of practical experience on their resumes. Additionally, the study emphasizes the considerable time and effort required from both professors and students. It also underscores the challenges associated with selecting an appropriate project for contribution, finding suitable tasks within the time constraints of a course, and engaging within an \ac{OSS} community. Similarly, through a qualitative analysis of semi-structured interviews with 24 faculty members, Postner et al. \cite{postner_impact_2019}
investigated how incorporating \ac{OSS} in education  influenced instructors' methodologies and perspectives regarding computing instruction, as well as the instructional resources deemed beneficial during the adoption of \ac{OSS} in educational settings. The study found that utilizing \ac{OSS} in teaching transformed instructors’ pedagogical approaches, enhanced instructors’ perspectives on the connections between academic content and professional practice, and has a positive impact on instructors’ attitudes and motivations when teaching. Moreover, the participants in the study emphasized that collaborative efforts among faculty, participation in \ac{OSS} education community gatherings, and the availability of shared learning materials have facilitated  adoption of \ac{OSS} in education.

Aiming to understand students' perspectives on learning through the utilization of \ac{OSS}, Pinto et al. conducted semi-structured interviews with 21 students who participated in courses integrating \ac{OSS} contributions, alongside an analysis of the commits contributed by students \cite{pinto_training_2019}.
The study unveiled that instructors employ diverse strategies for selecting \ac{OSS}, and students often require time to familiarize themselves with the chosen project before seeking tasks. Moreover, the study noted variations in the complexity and purposes of students' contributions. Participants in the study highlighted the benefits of engaging with real projects, such as skill enhancement, increased self-confidence, active contribution, and job offers. However, the study also identified several challenges faced by students, including technical challenges, engaging with the project community, and the short duration of the course.  Similarly, Nascimento et al. 
\cite{nascimento_does_2019}
conducted a  survey and interviews with  30 students who studied one of two software engineering courses that incorporated \ac{OSS} contribution. The study highlights the benefits of utilizing \ac{OSS}, which includes providing concrete examples and aid in understanding problems, aid in understanding theory, aid in consolidation and retention of contents, promotes active  participation and discussion, and confirms the applicability of theory.

Smith et al. \cite{smith_selecting_2014} delve into the challenges of identifying suitable \ac{OSS} projects for teaching software engineering, particularly, in the context of software maintenance and evolution. The authors underscore that despite the benefits of using \ac{OSS} projects to mimic real-world scenarios in teaching, the process of selecting appropriate projects can be both labor-intensive and complex, imposing a barrier to the broader integration of \ac{OSS} into software engineering and computing courses. In an effort to identify challenges faced by newcomers when selecting appropriate tasks to contribute to in \ac{OSS} projects, Santos et al. 
\cite{santos_hits_2022}
conducted a study involving 154 undergraduate students. In this study, students were tasked with reviewing issues from \ac{OSS} projects and determining the skills required to resolve them. The students' responses were then compared with the responses of six professional developers. The study's findings indicated that students often struggle in accurately identifying the necessary skills from issues in \ac{OSS} projects, with their responses frequently diverging from those of professionals. Particularly, students demonstrated a better understanding of skills related to databases, operating infrastructure, programming concepts, and programming languages, while their proficiency was comparatively lower in areas such as debugging and program comprehension.

 He et al. \cite{he_open_2023} reported their experience in teaching an \ac{OSS} onboarding course at Peking University. The course includes lectures, labs, and talks designed to equip students for effective participation in \ac{OSS} projects. During the course, students engage in semester-long projects, choosing either from recommended \ac{OSS} projects or their own preferences. The paper reports that 16 out of 19 students successfully made contributions to \ac{OSS}, with five of them continuing their involvement beyond the course. However, the outcomes varied notably based on the specific projects chosen.

Deng et al. \cite{deng_teaching_2020} proposed a software testing education approach that incorporates the contribution of \ac{OSS}. The approach includes five learning activities of differing complexity levels, all centered around real-world \ac{OSS} software projects. These activities are designed to enhance students' practical experience, foster networking opportunities, and boosting their confidence in their technical skills. The paper underscores the importance of collecting students' feedback for continuous course improvement. 

\section{Course Design} 
The software testing course in which the approach was experimented is
a four credit mandatory course for the Software Engineering undergraduate program. The course covers fundamental concepts in software testing covering various testing techniques, including white-box and black-box testing. It also delves into unit testing practices, automated test suites, and \ac{TDD}.  The course addresses integration, system, and acceptance testing and explores non-functional testing aspects. Below, we detail how the course was designed around the utilization of \ac{TBL} and \ac{OSS} contributions and collecting students' feedback.

\subsection{The Course Structure}
The course topics are organized into a total of six modules: a foundational module labeled as module zero, which introduces the course, and five \ac{TBL} modules. Below, we provide an overview of each module.

\subsubsection{Module zero} In this initial module, the instructor launches the course by providing a comprehensive overview. This includes introducing and assigning relevant readings about the \ac{TBL} method. Students are also requested to complete a profile survey, which helps in forming teams. Once the teams are established, a simulation of the readiness assurance phase in \ac{TBL} is conducted. This simulation revolves around the content of the course syllabus and the assigned \ac{TBL} readings as the basis for the test topics. Team-building activities are of paramount importance during this stage, aligning with the \ac{TBL} method's emphasis on fostering collaboration and teamwork.

\subsubsection{Modules one to five} This set of modules follows the typical 
 \ac{TBL} module activities as shown in Table \ref{tab:module}. The first three activities are more theory-focused, while the remaining two transition to practice-driven assignments towards the end.

Each module commences with students engaging with the assigned material, typically the course textbook. Supplementary resources such as videos, recorded classes, podcasts, and others may also be suggested. Following this preparatory phase, one class session is dedicated exclusively to the readiness assurance stage. This stage consists of an individual test followed by a team test. After the tests, students are given the opportunity to submit appeals for questions with which they may disagree with the instructor. In the subsequent class, the instructor provides a summary of the test results and conducts a review of the concepts that were found to be more challenging for the students.

In the practical segment of each module, two assignments are employed. Initially, all teams collaborate in-class on a common software testing problem. For instance, teams are tasked with generating test cases for a designated feature, utilizing a specified testing technique. Subsequently, students are required to apply the same testing technique to a feature within an \ac{OSS} project chosen by their team. While implementing the technique in the \ac{OSS}  project, team members can assist each other; however, each student is individually responsible for delivering tests for a specific feature. It's worth noting that in modules one and five, the \ac{OSS}  assignment is a collective effort by the team, resulting in a single submission.

\begin{table}
  \caption{Activities of each module}
  \label{tab:module}
  \begin{tabular}{lll}
    \toprule
    \# & Focus & Activity\\
    \midrule
    1 & Theory & Individual study with assigned material\\
    2 & Theory & Individual and team-based test\\
    3 & Theory & Appeals and oral-feedback from the instructor\\
    4 & Practice & In-class application exercise\\
    5 & Practice & \ac{OSS} application project\\
  \bottomrule
\end{tabular}
\end{table}

\subsection{\ac{TBL} Modules}

Table \ref{tab:structure} presents the course schedule, including the covered topics and activities in each module.

\begin{table*}
  \centering
  \caption{Course schedule and covered topics}
  \label{tab:structure}
    \begin{tabular}{cp{3cm}p{7.8cm}p{2cm}}
    \toprule
    \textbf{Week} & \multicolumn{1}{c}{\textbf{Module}} & \multicolumn{1}{c}{\textbf{Activities}} & \textbf{Assignment/ Grading} \\
    \midrule

    1     & \multirow{2}{3cm}{0 - Overview} & \multicolumn{1}{l}{Course overview, syllabus, team formation} &  \\
    2     & \multicolumn{1}{l}{} & \multicolumn{1}{l}{Individual preparation, \ac{TBL}simulation} &  \\

    \midrule

    3     & \multirow{3}{3cm}{1 - Test Strategy} & \multicolumn{1}{l}{Individual study, Individual and Team Readiness Assurance Tests (iRAT and tRAT)} & iRAT1, tRAT1 \\
    4     & \multicolumn{1}{l}{} & \multicolumn{1}{l}{Review class with instructor feedback, In-class application exercise (AE)} & AE1 \\
    5     & \multicolumn{1}{l}{} & \multicolumn{1}{l}{Presentation of \ac{OSS} project options} & \ac{OSS} 1 \\

    \midrule
    
    6     & \multirow{2}{3cm}{2 - Functional Testing / System Tests} 
        & \multicolumn{1}{l}{Individual study, Individual and Team Readiness Assurance Tests (iRAT and tRAT)} & iRAT2, tRAT2 \\
    7     & \multicolumn{1}{l}{} & \multicolumn{1}{l}{Review class with instructor feedback, In-class application exercise (AE)} & AE2, \ac{OSS} 2 \\

    \midrule
    
    8     & Feedback & Midterm peer evaluation and course retrospective &  \\
    
    \midrule
    
    9     & \multirow{3}{3cm}{3 - Structural Testing / Unit and Integration Tests} & \multicolumn{1}{l}{Individual study, Individual and Team Readiness Assurance Tests (iRAT and tRAT)} & iRAT3, tRAT3 \\
    10    & \multicolumn{1}{l}{} & \multicolumn{1}{l}{Review class with instructor feedback, In-class application exercise (AE)} & AE3, \ac{OSS} 3 \\
    & \multicolumn{1}{l}{} &  & \\

    \midrule
    
    11    & \multirow{2}{3cm}{4 - TDD} & \multicolumn{1}{l}{Individual study, Individual and Team Readiness Assurance Tests (iRAT and tRAT)} & iRAT4, tRAT4 \\
    12    & \multicolumn{1}{l}{} & \multicolumn{1}{l}{Review class with instructor feedback, In-class application exercise (AE)} & AE4, \ac{OSS} 4 \\

    \midrule
    
    13    & \multirow{2}{3cm}{5 - Non-functional Testing} & \multicolumn{1}{l}{Individual study, Individual and Team Readiness Assurance Tests (iRAT and tRAT)} & iRAT5, tRAT5 \\
    14    & \multicolumn{1}{l}{} & \multicolumn{1}{l}{Review class with instructor feedback, In-class application exercise (AE)} & AE5, \ac{OSS} 5 \\

    \midrule
    
    15    & Feedback & Final term peer evaluation and course retrospective & PR \\

    \bottomrule
    \end{tabular}%

\end{table*}%

\subsubsection{Module 1} This module covers essential content related to testing concepts required for crafting a test strategy in a software project. Students explore topics such as the test pyramid, testing levels, various testing techniques, and the roles involved in the testing process. 

In the application exercise for this module (AE1), students are tasked with proposing a testing strategy for a provided software development case. Following this, in the \ac{OSS}1 project assignment, they analyze the existing testing strategy of their chosen \ac{OSS} project and have the opportunity to make recommendations for its enhancement. \ac{OSS}1 allows students to immerse themselves in the \ac{OSS} project, gaining insights into the types of tests and automation tools in use, available documentation, and contribution guidelines.

At this stage, students are strongly encouraged to fork the project into their accounts and set up their development environment. This preparatory step ensures that they become acquainted with the project and are well-prepared to make their initial contributions in the upcoming module.

\subsubsection{Module 2} This module is dedicated to instructing students in functional (black-box) testing techniques. In the in-class application exercise (AE2), students are tasked with creating test cases for a proposed feature, using techniques such as equivalence partitioning and boundary value analysis. This same activity is mirrored in \ac{OSS}2, where each student selects a specific feature to generate test cases, subsequently executing these tests within the development environment and reporting the outcomes. In case of a test failure, students are encouraged to submit a bug report to the project community, following the established issue tracking system. Successful acceptance of their bug report makes them eligible for extra points in this activity. Additionally, in this module, an optional extra points activity is available, focusing on automating system tests using automation tools like Selenium\footnote{https://www.selenium.dev}.

\subsubsection{Module 3} This module focuses on structural (white-box) testing, unit testing, and integration testing. In the in-class application exercise (AE3), teams are given the task of generating test cases for a specific method, utilizing multiple-condition coverage criteria (MCC). In the \ac{OSS}3 assignment, each student selects a method based on criteria defined by the instructor and formulates test cases using the MCC approach.

Following the learning activities proposed by Deng et. al. \cite{deng_teaching_2020}, this assignment requires students to execute the existing unit test cases, analyze the testing coverage, identify areas with inadequate coverage, and then develop new test cases using the MCC criteria. Subsequently, they reassess the testing coverage and have the option to submit a pull request to the project's community. Students can earn extra points if their pull request is accepted.

\subsubsection{Module 4} Module 4 delves into the \ac{TDD} method. In the application exercise (AE4), the instructor assigns a programming task to teams, guiding them through the implementation process using the TDD approach. Subsequently, students put TDD into practice in \ac{OSS}4, following the learning activity defined in Deng et. al. \cite{deng_teaching_2020}. In this context, students are required to select an unimplemented feature and develop it following the TDD cycle. They have the opportunity to initiate a pull request for the implemented feature, and if it is accepted, they become eligible for extra points.

\subsubsection{Module 5} Module 5 is dedicated to providing an overview of non-functional testing, with a particular emphasis on security testing. In this module, the application exercise (AE5) involves performing static analysis security testing for the \ac{OSS} project, with the results obtained during the in-class session.

Subsequently, in \ac{OSS}5, each student is tasked with selecting three vulnerabilities or security hot spots for analysis. They are required to classify these findings as true or false positives and, in the case of a legitimate security issue, report it to the project community while also providing potential solutions. Once again, the possibility of earning extra points is available if a security issue report is accepted by the project community.

\subsection{The \ac{OSS}   Practice} \label{OSSpractice}
Though the selection of appropriate \ac{OSS} projects plays a crucial role in the successful integration of \ac{OSS} contributions into software engineering education, it can indeed be a challenging task given the vast number of repositories available on platforms like GitHub. Moreover, as emphasized by Deng et. al. \cite{deng_teaching_2020}, not every \ac{OSS} project is suitable for students to gain valuable insights into software testing.

Therefore, in this course, we adopt a hybrid strategy for project selection, as defined by Deng et. al. \cite{deng_teaching_2020}. This approach enables instructors to curate a list of candidate \ac{OSS} projects for the entire class. Subsequently, students can make their selections from these candidates, taking into account their personal preferences, backgrounds, and levels of experience. We adapted the criteria used by Smith et. al. \cite{smith_selecting_2014} and Deng et. al. \cite{deng_teaching_2020} for the selection of the projects:

\begin{enumerate}
    \item The project must be actively developed and have multiple active committers.
    \item The project's size should be neither too small nor too large.
    \item The programming language used in the project should have been taught in previous courses to students.
    \item The application domain of the project should be appealing to students and relatively easy for them to learn about.
    \item The project should be well-documented, enabling students to access information about testing practices and understand project features.
    \item The project should have a set of implemented unit tests.
    \item The development environment for the project should be relatively easy to set up.
\end{enumerate}

With this set of criteria in mind, identifying suitable candidate projects can indeed be a challenge. To address this, students are encouraged to actively participate in the search for projects that align with these criteria. Additionally, projects developed within the university's laboratory settings can be excellent options, as they offer the advantage of direct instructor contact with the development team, facilitating coordination and support. This collaborative approach can help ensure the selection of projects that meet the specific needs and objectives of the course.

During the four semesters covered in this paper, we primarily opted for Java-based projects, considering that the majority of students possessed prior experience with this programming language from previous courses. The selected projects included both standalone Java applications and those integrating Java in the back-end coupled with JavaScript in the front-end.

\subsection{Peer Evaluation and Course Retrospective}

Peer evaluations play a critical role in the \ac{TBL} approach, as emphasized by Michaelsen and Sweet \cite{michaelsen_essential_2008}. In this course, these evaluations occur both midway through and at the conclusion of the course. 

During the midterm evaluation, the focus is on qualitative feedback, with the intention of enabling team members to offer and receive constructive feedback from their peers. The instructor emphasizes to students how this practice is valuable not only in an educational context but also in a professional work environment, making it a valuable skill to develop.

Towards the end of the semester, students are tasked with providing quantitative evaluations of their peers. The resulting assessments are used to adjust the grading of team-based work over the course of the semester. In this evaluation process, each student receives a fixed number of points to distribute among their team members. If they perceive that everyone contributed equally to team assignments, they distribute the points evenly. Conversely, if they believe that one team member contributed more while another contributed less, they adjust the points accordingly.

It's worth noting that these evaluations are conducted anonymously, with all team members evaluating each other, in addition to conducting a self-evaluation. Tools such asTeamMates\footnote{https://teammatesv4.appspot.com/} are employed to facilitate the distribution of questionnaires and the confidential reporting of results to each student.

In addition to peer evaluations, course retrospectives are conducted both at the midpoint and the conclusion of the semester. This concept is inspired by agile retrospectives \cite{schon_shift_2023} and serves the purpose of gathering feedback from students regarding the course itself. Students are prompted to identify what is working effectively and should be maintained, pinpointing areas that are not working well and should be reevaluated, suggesting potential improvements, and proposing ideas for future sessions in the course. These retrospectives offer opportunities for making real-time adjustments and improvements during the current semester and for enhancing the course in subsequent semesters \cite{krehbiel_agile_2017}.

\subsection{Grading}

An effective grading system for \ac{TBL} should incentives both individual contributions and successful team collaboration, while also addressing equity concerns that often arise when group work factors into an individual's grade \cite{michaelsen_essential_2008}.

Thus, we opted to grade all \ac{TBL} activities and introduce a multiplier factor, known as the Fink method \cite{cestone_peer_2008}, obtained through the peer evaluations to adjust the team-based tasks. The final grade for the course is calculated according to Equation \ref{eq:grade}:

\begin{multline} \label{eq:grade}
Final Grade = iRAT*W1 + tRAT*PE*W2 \\ + AE*PE*W3 + OSS*50\%
\end{multline}

where:

\begin{conditions*}
iRAT & average of the grades obtained in the individual readiness assurance tests \\
tRAT & average of the grades obtained in the team readiness assurance tests \\
AE & average of the scores obtained in in-class team 
        application exercises \\
W1, W2, W3 & weights to be defined by the class in the 
        second week of the course \\ 
PE & the multiplier factor generated from peer evaluations 
        at the end of the course \\
\ac{OSS}   & the weighted average of \ac{OSS}   deliveries, calculated 
        according to equation \ref{eq:OSS} \\
\end{conditions*}
  
\begin{multline}  \label{eq:OSS}
OSS = OSS1*15\% + OSS2*25\% + OSS3*25\% + \\ OSS4*20\% + OSS5*15\%
\end{multline}

\section{Course Evaluation} 
The aforementioned designed course was implemented 
 since the beginning of the second semester in 2021 until the first semester of 2023. 

Table \ref{tab:sections} displays sections data pertaining to this course over the specified period. A total of seven course sections were offered, encompassing two sections with online classes, due to the pandemics, and five sections with in-person classes, starting from 2022. The course was completed for 318 students; 165 of these students, constituting 52\% of the overall cohort, voluntarily participated in a survey aimed at assessing the course's effectiveness.

\begin{table}
  \centering
  \caption{Software Testing course sections that applied the approach} \label{tab:sections}
    \begin{tabular}{lrlrr}
    \toprule
    Semester & \multicolumn{1}{l}{Section}  & Mode  & \multicolumn{1}{l}{Students} & \multicolumn{1}{l}{Survey Participants} \\
    \midrule
    21.2  & 1     & Online & 54    & 45 \\
    21.2  & 2     & Online & 29    & 27 \\
    22.1  & 1     & In-Person & 41    & 39 \\
    22.2  & 1     & In-Person & 21    & 5 \\
    22.2  & 2     & In-Person & 62    & 19 \\
    23.1  & 1     & In-Person & 56    & 13 \\
    23.1  & 2     & In-Person & 55    & 17 \\
    \midrule
        Total & & & 318 & 165 \\
    \bottomrule
    \end{tabular}%
  \label{tab:addlabel}%
\end{table}%

Towards the conclusion of each semester, spanning from 21.2 to 23.1, as presented in the table, students were invited to participate in a survey, focusing on aspects related to the course's design.
In our endeavor to facilitate comparative analysis with previous research, we structured our questionnaire based on a prior survey conducted in two \ac{TBL}-based courses, as documented by Matalonga et. al. \cite{matalonga_deploying_2017}, which itself drew inspiration from surveys featured in the book by Michaelsen et. al. \cite{michaelsen_team-based_2002}.

The original surveys outlined in the book comprised a total of 11 questions, with Matalonga et. al. \cite{matalonga_deploying_2017} subsequently incorporating seven additional questions aimed at evaluating deployment-related facets. In our adaptation of the questionnaire, we omitted two questions concerning the use of traditional \ac{TBL}IF AT forms, as our course utilized online questionnaires instead. Additionally, we excluded one question concerning the frequency of attendance in application-exercise classes, as it was deemed inapplicable within our course context. Furthermore, one question inquiring about the impact of large class size was also removed because our course was essentially on par with other courses within the program.
In lieu of these exclusions, we introduced one question (Q7) regarding the contribution of the in-class application exercise (AE) and three new questions designed to elicit student perceptions regarding \ac{OSS} practice. The latter three questions were added commencing in the 22.2 semester.


Table \ref{tab:results} presents the survey's questions and results.

\begin{table*}
  \centering
  \caption{Students' perception on the course teaching methodology}
    \begin{tabular}{rp{12cm}rrrrr}
    \toprule
    \multicolumn{1}{l}{\#} & Question & \multicolumn{1}{l}{\% 1-2} & \multicolumn{1}{l}{\% 3-5} & \multicolumn{1}{l}{\% 6-7} & \multicolumn{1}{l}{Avg} & \multicolumn{1}{l}{Sd} \\
    \midrule
    1     & Do you find that a team learning strategy has more or fewer advantages than the traditional "lecture, mid-term and final exams" approach? (1=Fewer advantages; 7=More)  & 6\%   & 22\%  & 72\%  & 5.90  & 1.51 \\
    2     & How much have you learned with \ac{TBL}compared to a traditional class (1=Less, 7=More)  & 4\%   & 39\%  & 56\%  & 5.52  & 1.39 \\
    3     & How much have you studied with \ac{TBL}compared to a traditional class (1=Less, 7=More)  & 7\%   & 47\%  & 46\%  & 5.19  & 1.57 \\
    4     & How stressful did you perceive the individual RATs to be, in this class? (1=Very stressful; 7=Not stressful at all)  & 8\%   & 50\%  & 42\%  & 4.91  & 1.62 \\
    5     & How stressful did you perceive the team RATs to be, in this class? (1=Very stressful; 7=Not stressful at all)  & 10\%  & 30\%  & 60\%  & 5.33  & 1.75 \\
    6     & Do you think the "appeal" procedure is useful? (1=Not at all, 7=Very)  & 10\%  & 33\%  & 57\%  & 5.38  & 1.78 \\
    7     & Were the [online synchronous | in-person] classes of application activities helpful? (1=Not at all; 7=Very) & 12\%  & 35\%  & 54\%  & 5.25  & 1.81 \\
    8     & How rewarding was to apply the theoretical knowledge to application-exercise classes? (1=Not at all, 7=Very rewarding)  & 7\%   & 35\%  & 58\%  & 5.53  & 1.58 \\
    9     & How compatible is team learning with your personal learning style (in other words, does a learn learning strategy match your way of learning)? (1=Not compatible at all; 7=Extremely compatible)  & 10\%  & 53\%  & 36\%  & 4.81  & 1.68 \\
    10    & How likely are you to keep some of the friendships you made made in your group, outside of this class? (1=Not likely at all; 7=Extremely likely)  & 11\%  & 47\%  & 42\%  & 5.03  & 1.76 \\
    11    & In terms of managing day-to-day activities in class (such as handing out tests, grades, etc.), how complicated did you think team learning was? (1=Extremely complicated; 7=Not complicated at all)  & 5\%   & 52\%  & 43\%  & 5.02  & 1.50 \\
    12    & How valuable do you think a team learning approach would be for some of your other classes? (1=Not valuable at all; 7=Extremely valuable)  & 5\%   & 32\%  & 62\%  & 5.69  & 1.50 \\
    13    & How satisfied were you with the amount of learning you gained in this class? (1=Not at all satisfied; 7=Extremely satisfied)  & 4\%   & 41\%  & 56\%  & 5.47  & 1.41 \\
    14    & Overall, how would you rate your experience in this class? (1=Terrible; 7= Excellent)  & 4\%   & 38\%  & 59\%  & 5.54  & 1.30 \\
    15    & How likely are you to recommend this class to other students? (1=Not at all likely; 7=Extremely likely)  & 4\%   & 19\%  & 77\%  & 6.12  & 1.38 \\
    16    & How difficult was it to carry out the test activities in the \ac{OSS} ? (1=Very easy; 7=Extremely difficult) & 4\%   & 30\%  & 67\%  & 5.65  & 1.42 \\
    17    & How much did the activities related to the \ac{OSS}  project contribute to your learning? (1=Not at all; 7=Very) & 4\%   & 50\%  & 46\%  & 5.11  & 1.45 \\
    18    & How satisfied were you with the opportunity to contribute to an OSS project? (1=Not at all; 7=Extremely satisfied) & 19\%  & 59\%  & 22\%  & 4.15  & 1.70 \\
    \bottomrule
    \end{tabular}%
  \label{tab:results}%
\end{table*}%


 Consistent with the methodologies employed in the aforementioned studies, we adopted a uniform rating scale, which ranged from 1 to 7 points. Subsequently, following a similar approach, we analyzed the responses, categorizing them into three distinct subranges as indicated by the "\% n-m" columns in Table \ref{tab:results}: low scores (1 and 2 points), intermediate scores (3 to 5 points), and high scores (6 and 7 points). Furthermore, we determined the average score for each question ("Avg" column) and calculated the standard deviation ("Sd" column).

 We observe that questions related to the \ac{TBL} approach and its application within the course received consistently high scores. Specifically, questions 1, 2, and 12 yielded notably positive responses. According to students' perceptions, \ac{TBL} demonstrated a range of advantages over traditional teaching methods, with 72\% of respondents assigning high scores to this question. Furthermore, the majority of students reported enhanced learning outcomes compared to traditional approaches, with 56\% awarding high scores. 
 Additionally, 62\% of students expressed a belief that the \ac{TBL} could prove to be valuable in other courses. 
 However, when comparing how much they studied with \ac{TBL} with a traditional class (Q3), 47\% of respondents reported intermediate scores, while 46\% of them reported high scores.

In accordance with questions 14 and 15, students held their course experience in high regard, with 59\% of participants assigning high scores. Moreover, 77\% of respondents indicated a strong likelihood of recommending the course to their peers.

In reference to the \ac{RAT}, it is notable that students found team \ac{RAT}s to be less stressful, as evidenced by 60\% of participants assigning high scores in this regard (Q4). In contrast, individual \ac{RAT}s garnered comparatively lower satisfaction, with 42\% of respondents providing high scores (Q5). Additionally, the "appeal" procedures, designed to address concerns and disputes, were deemed useful by 57\% of the respondents, indicating their appreciation for this aspect of the testing process (Q6).

In the context of in-class application exercises (Q7 and Q8), students expressed a positive sentiment, with 54\% of them providing high scores to signify their appreciation for the utility of these exercises. Furthermore, a substantial 58\% of respondents indicated that they found the process of applying theoretical knowledge to these application-exercise classes to be rewarding. This observation underscores a smooth and effective transition from theory-related activities, such as assigned readings, readiness assurance tests, appeals, and instructor feedback, to practical, hands-on exercises.

The majority of students (53\%) reported intermediate scores when assessing the compatibility of their learning style with \ac{TBL} (Q9). A similar distribution was observed for the question regarding the complexity of managing day-to-day activities (Q11), with 52\% of respondents assigning intermediate scores. These intermediate results are likely attributable to the substantial volume of activities demanded by the \ac{TBL}  approach over the course of the semester, especially when compared to the more traditional "lecture, mid-term, and final exams" model.
Questions 16 to 18 were designed to gauge students' perceptions of their involvement in contributing to tests in \ac{OSS} projects during the course. The data reveals that a significant portion of students encountered challenges in executing the test assignments within the \ac{OSS} projects, with 67\% of high scores. However, when considering how much these activities contributed to their learning, the responses varied, with 50\% assigning intermediate scores and 46\% assigning high scores. In a similar vein, 59\% reported intermediate scores regarding their satisfaction with the opportunity to contribute to \ac{OSS} projects.

It's noteworthy that the lower satisfaction levels regarding the opportunity to contribute to \ac{OSS} projects may be influenced by the difficulties faced during the process. It is important to highlight that for these questions, we received only 54 responses as they were introduced starting from the 22.2 semester.

\begin{figure*}
  \centering
  \includegraphics[width=\linewidth]{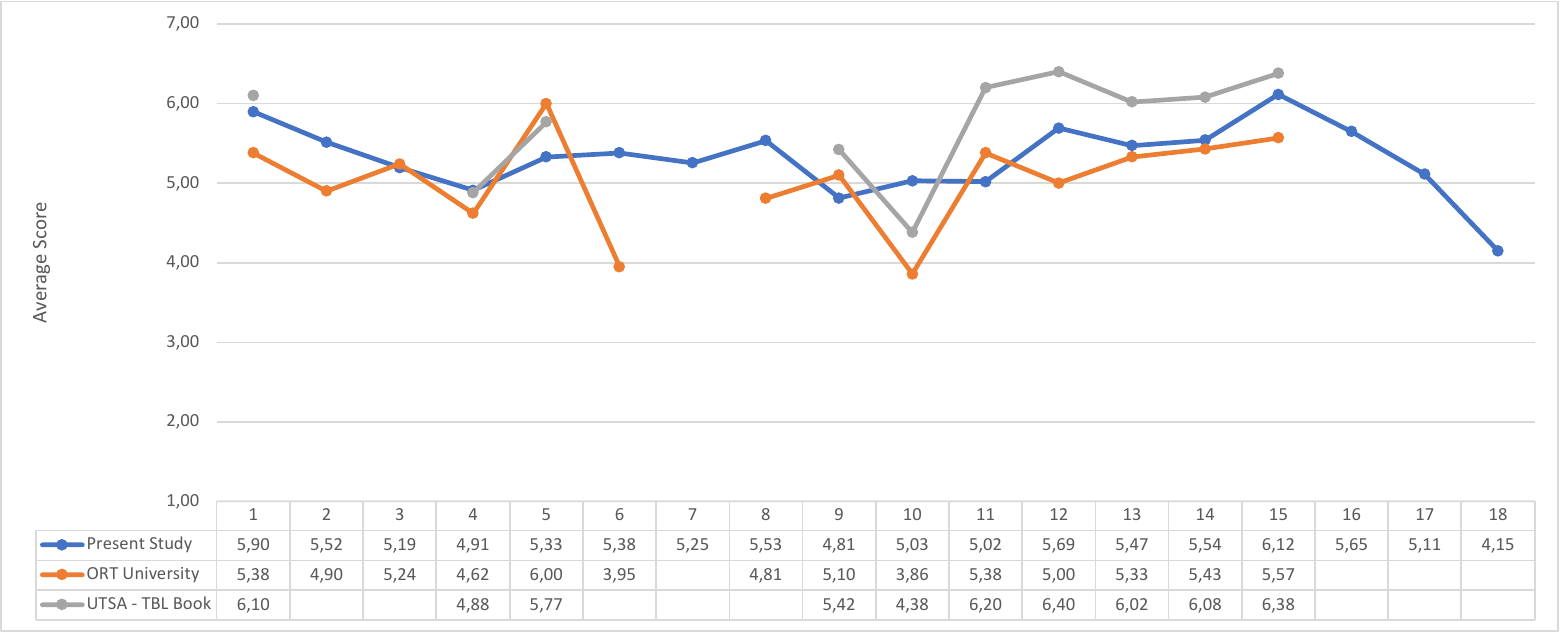}
  \caption{Comparison of the average scores per question} \label{fig:comparison}
  \Description{A line graph displaying the comparison of scores between the Software Testing course and the results obtained in \cite{matalonga_deploying_2017} and \cite{michaelsen_team-based_2002}}.
\end{figure*}

Figure \ref{fig:comparison} presents a comparative analysis of the average scores for each question across multiple surveys. This includes our Software Testing course (present study), the average scores at Universidad ORT Uruguay (ORT) as documented in Matalonga et. al. \cite{matalonga_deploying_2017}, and the results obtained from students at the University of Texas at San Antonio (UTSA) as published in the \ac{TBL} book by Michaelsen et. al. \cite{michaelsen_team-based_2002}. The empty points on the chart represent questions that were not included in the survey.

An observation from the chart is that the trend in the average score values across the three surveys demonstrates remarkable similarity. The present study values typically fall between those of UTSA and ORT, with one notable exception being question 10. In this question, students at our university appear more inclined to maintain friendships with their team members outside of class.

Additionally, the present study average scores surpass those of ORT in questions 2 (regarding learning with \ac{TBL} compared to traditional classes), 6 (concerning the utility of the appeal procedure), and 8 (pertaining to the reward of applying theoretical knowledge in practical exercises). Notably, UTSA did not include these specific questions in their survey.

As depicted in Figure \ref{fig:comparison}, questions 9 (regarding the compatibility of \ac{TBL}  with the student's learning style) and 11 (concerning the challenges in managing day-to-day activities in class) exhibited relatively lower scores in our course when compared to ORT and UTSA. This discrepancy might be attributed to the transition that occurred during the period of our analysis, where students shifted from online classes back to in-person classes. Many students reported difficulties in readapting to this change in the learning format, which could account for the differences in their responses to these particular questions.

Moreover, it is noteworthy that question 18 falls very close to the neutral score, representing the lowest score among all the questions. This observation underscores the fact that, although students generally favor the \ac{TBL}  approach, they express dissatisfaction with contributing to an \ac{OSS} project.

\section{Reflections and Recommendations}

\textit{\ac{OSS} First Contribution.}
Our course is typically taken by students at the outset of their third year in the undergraduate Software Engineering program. For the majority, around 95\%, this marks their first experience with contributing to an \ac{OSS} project. Many students encounter challenges when setting up the development environment on their personal computers, primarily due to the intricate dependencies involved in certain projects. Consequently, it becomes crucial to meticulously select \ac{OSS} projects for student engagement. Beyond the criteria outlined in Section \ref{OSSpractice}, it is recommended to assess the complexity of configuring the development environment before assigning projects to students. Additionally, the quality of repository documentation plays a pivotal role as it facilitates the learning process for students venturing into \ac{OSS} contributions.

\textit{\ac{OSS} assignments grading.}
In larger classrooms, the grading of \ac{OSS} testing assignment reports, particularly in modules two, three, and four of our course, can become a substantial task due to the high volume of tasks. To streamline the grading process, we suggest the use of structured templates for the students' reports. Additionally, establishing objective rubrics can help ensure consistency in grading. Involving graduate students as graders can also aid in efficiently managing the grading workload.

\textit{\ac{OSS} good first issue.}
Numerous \ac{OSS} projects on GitHub employ the "good first issue" label to identify tasks suitable for beginners. These particular issues serve as excellent options for the \ac{TDD} assignment. Nevertheless, it is important to acknowledge that not all projects may have a sufficient number of these beginner-friendly tasks to accommodate every student. In such instances, students may be encouraged to suggest new, smaller features or even utilize \ac{TDD} to address and fix software bugs.

\textit{Use of tools}.
Given the substantial workload associated with the course activities, the utilization of appropriate tools is imperative. We have found success in employing TeamMates\footnote{https://teammatesv4.appspot.com/} for peer evaluations. TeamMates is an \ac{OSS} project\footnote{https://github.com/TEAMMATES/teammates} designed for feedback management in education. This platform offers user-friendly features that simplify the enrollment of students and the setup of team peer-review sessions. Notably, TeamMates includes unique question types, such as those allowing respondents to allocate points among their team members based on individual contributions to team-based tasks.

Due to the considerable time required for manual grading of the traditional \ac{TBL} scratch card, we have transitioned to using Moodle's questionnaire feature for digital readiness assurance tests, despite encountering certain limitations. Regrettably, Moodle does not provide comprehensive support for readiness assurance tests linked with \ac{TBL}, especially those requiring group responses in questionnaires. Nevertheless, it proves invaluable for automating the scoring process of reports submitted for in-class team application exercises and \ac{OSS} deliveries, thereby streamlining the assessment procedure.

\textit{Extra points for pull requests}.
In line with the study by Deng et. al. \cite{deng_teaching_2020}, we do not expect that the pull requests submitted by students will necessarily be approved or receive prompt responses. The review and acceptance process for pull requests can vary in duration, contingent upon the specific \ac{OSS} project and the workload of project maintainers. To encourage students to engage in pull requests submissions, we introduced a system that awarded extra points for both the submission and, subsequently, the acceptance of pull requests. However, we soon observed that this incentive prompted students to submit pull requests lacking genuine contributions, which inundated the \ac{OSS} community with superfluous requests.

As a response, we opted to discontinue offering extra points for pull requests submissions, while retaining the incentive for pull requests acceptance. Additionally, we began to emphasize the significance of volunteers' work in free and \ac{OSS} projects, imparting a sense of responsibility to students. The aim is to encourage them to only submit pull requests when they genuinely contribute to the projects, thus maintaining the integrity and efficiency of the \ac{OSS} community.

\textit{Classroom Layout.}
One of the challenges encountered when implementing \ac{TBL} is that the majority of classrooms are primarily designed for traditional lectures, making it less conducive to fostering meaningful interaction among students. Recent studies highlight how the classroom layout impacts interaction and learning \cite{asino_student_2019, rands_room_2017}. In our current context, all classrooms maintain a traditional layout, sometimes requiring students to stand for team readiness assurance test discussions or use their laptops on their laps. We expect that in the near future, active learning classrooms will be available and, hopefully, will enhance the learning experience and outcomes for our students.

\section{Conclusion} 

This paper discusses the importance of evolving beyond traditional lecture-based teaching in undergraduate Software Engineering programs, emphasizing the active engagement of students in practical, iterative development. The study outlines an approach that gradually transitions students from theory to practice, starting with \ac{TBL} and culminating in active contributions to \ac{OSS} projects. This methodology was applied over four semesters in a Software Testing course, involving a significant number of students. 

In general, the responses obtained from our survey align with the findings of previous experience reports. Students clearly perceive that the Team-Based Learning (TBL) approach offers more advantages compared to a traditional lecture-oriented course. They value the in-class application exercises and find the application of theoretical knowledge in these activities to be rewarding. 

Overall, this pedagogical model seeks to bridge the gap between theoretical knowledge and real-world practice in Software Testing, by combining TBL and OSS contribution, and therefore aligning with the broader goals of Software Engineering education. However, it is evident that students encounter challenges when facing their first contribution to an \ac{OSS}  projects. 

To address this issue, we recognize the need for greater support and guidance to help students overcome the challenges associated with their first \ac{OSS}  contribution. In response, we are planning to introduce activities that will familiarize students with the procedures and requirements of \ac{OSS}  contributions, thereby enhancing their preparedness for this task. We will also improve our end-of-semester survey to better understand students' perspectives on satisfaction related to OSS contribution.  

Additional research on this pedagogical approach could also delve into the content of the pull-requests and assess the impact of students' testing contributions to the chosen Open Source Software (OSS) repositories.

\section{Acknowledgments}

Special thanks to the students who actively participated in the survey, significantly contributing to this experience report and enhancing our teaching practices.


\bibliographystyle{ACM-Reference-Format}
\bibliography{references}


\begin{thebibliography}{26}


\ifx \showCODEN    \undefined \def \showCODEN     #1{\unskip}     \fi
\ifx \showDOI      \undefined \def \showDOI       #1{#1}\fi
\ifx \showISBNx    \undefined \def \showISBNx     #1{\unskip}     \fi
\ifx \showISBNxiii \undefined \def \showISBNxiii  #1{\unskip}     \fi
\ifx \showISSN     \undefined \def \showISSN      #1{\unskip}     \fi
\ifx \showLCCN     \undefined \def \showLCCN      #1{\unskip}     \fi
\ifx \shownote     \undefined \def \shownote      #1{#1}          \fi
\ifx \showarticletitle \undefined \def \showarticletitle #1{#1}   \fi
\ifx \showURL      \undefined \def \showURL       {\relax}        \fi
\providecommand\bibfield[2]{#2}
\providecommand\bibinfo[2]{#2}
\providecommand\natexlab[1]{#1}
\providecommand\showeprint[2][]{arXiv:#2}

\bibitem[Asino and Pulay(2019)]%
        {asino_student_2019}
\bibfield{author}{\bibinfo{person}{Tutaleni~I. Asino} {and} \bibinfo{person}{Alana Pulay}.} \bibinfo{year}{2019}\natexlab{}.
\newblock \showarticletitle{Student {Perceptions} on the {Role} of the {Classroom} {Environment} on {Computer} {Supported} {Collaborative} {Learning}}.
\newblock \bibinfo{journal}{\emph{TechTrends}} \bibinfo{volume}{63}, \bibinfo{number}{2} (\bibinfo{date}{March} \bibinfo{year}{2019}), \bibinfo{pages}{179--187}.
\newblock
\showISSN{1559-7075}
\urldef\tempurl%
\url{https://doi.org/10.1007/s11528-018-0353-y}
\showDOI{\tempurl}


\bibitem[Awatramani and Rover(2015)]%
        {awatramani_team-based_2015}
\bibfield{author}{\bibinfo{person}{Mihir Awatramani} {and} \bibinfo{person}{Diane Rover}.} \bibinfo{year}{2015}\natexlab{}.
\newblock \showarticletitle{Team-based learning course design and assessment in computer engineering}. In \bibinfo{booktitle}{\emph{2015 {IEEE} {Frontiers} in {Education} {Conference} ({FIE})}}. \bibinfo{pages}{1--9}.
\newblock
\urldef\tempurl%
\url{https://doi.org/10.1109/FIE.2015.7344227}
\showDOI{\tempurl}


\bibitem[Brito et~al\mbox{.}(2018)]%
        {brito_floss_2018}
\bibfield{author}{\bibinfo{person}{Moara~Sousa Brito}, \bibinfo{person}{Fernanda~Gomes Silva}, \bibinfo{person}{Christina von~Flach G.~Chavez}, \bibinfo{person}{Debora~C. Nascimento}, {and} \bibinfo{person}{Roberto~A. Bittencourt}.} \bibinfo{year}{2018}\natexlab{}.
\newblock \showarticletitle{{FLOSS} in software engineering education: an update of a systematic mapping study}. In \bibinfo{booktitle}{\emph{Proceedings of the {XXXII} {Brazilian} {Symposium} on {Software} {Engineering}}} \emph{(\bibinfo{series}{{SBES} '18})}. \bibinfo{publisher}{Association for Computing Machinery}, \bibinfo{address}{New York, NY, USA}, \bibinfo{pages}{250--259}.
\newblock
\showISBNx{978-1-4503-6503-1}
\urldef\tempurl%
\url{https://doi.org/10.1145/3266237.3266249}
\showDOI{\tempurl}


\bibitem[Cestone et~al\mbox{.}(2008)]%
        {cestone_peer_2008}
\bibfield{author}{\bibinfo{person}{Christina~M. Cestone}, \bibinfo{person}{Ruth~E. Levine}, {and} \bibinfo{person}{Derek~R. Lane}.} \bibinfo{year}{2008}\natexlab{}.
\newblock \showarticletitle{Peer assessment and evaluation in team-based learning}.
\newblock \bibinfo{journal}{\emph{New Directions for Teaching and Learning}} \bibinfo{volume}{2008}, \bibinfo{number}{116} (\bibinfo{date}{Dec.} \bibinfo{year}{2008}), \bibinfo{pages}{69--78}.
\newblock
\showISSN{0271-0633}
\urldef\tempurl%
\url{https://doi.org/10.1002/tl.334}
\showDOI{\tempurl}
\newblock
\shownote{Publisher: John Wiley \& Sons, Ltd}.


\bibitem[Deng et~al\mbox{.}(2020)]%
        {deng_teaching_2020}
\bibfield{author}{\bibinfo{person}{Lin Deng}, \bibinfo{person}{Josh Dehlinger}, {and} \bibinfo{person}{Suranjan Chakraborty}.} \bibinfo{year}{2020}\natexlab{}.
\newblock \showarticletitle{Teaching {Software} {Testing} with {Free} and {Open} {Source} {Software}}. In \bibinfo{booktitle}{\emph{2020 {IEEE} {International} {Conference} on {Software} {Testing}, {Verification} and {Validation} {Workshops} ({ICSTW})}}. \bibinfo{pages}{412--418}.
\newblock
\urldef\tempurl%
\url{https://doi.org/10.1109/ICSTW50294.2020.00074}
\showDOI{\tempurl}


\bibitem[He et~al\mbox{.}(2023)]%
        {he_open_2023}
\bibfield{author}{\bibinfo{person}{Hao He}, \bibinfo{person}{Minghui Zhou}, \bibinfo{person}{Qingye Wang}, {and} \bibinfo{person}{Jingyue Li}.} \bibinfo{year}{2023}\natexlab{}.
\newblock \showarticletitle{Open {Source} {Software} {Onboarding} as a {University} {Course}: {An} {Experience} {Report}}. In \bibinfo{booktitle}{\emph{2023 {IEEE}/{ACM} 45th {International} {Conference} on {Software} {Engineering}: {Software} {Engineering} {Education} and {Training} ({ICSE}-{SEET})}}. \bibinfo{pages}{324--336}.
\newblock
\urldef\tempurl%
\url{https://doi.org/10.1109/ICSE-SEET58685.2023.00037}
\showDOI{\tempurl}
\newblock
\shownote{ISSN: 2832-7578}.


\bibitem[{Joint Task Force on Computing Curricula IEEE Computer Society Association for Computing Machinery}(2014)]%
        {joint_task_force_on_computing_curricula_ieee_computer_society_association_for_computing_machinery_curriculum_2014}
\bibfield{author}{\bibinfo{person}{{Joint Task Force on Computing Curricula IEEE Computer Society Association for Computing Machinery}}.} \bibinfo{year}{2014}\natexlab{}.
\newblock \bibinfo{title}{Curriculum {Guidelines} for {Undergraduate} {Degree} {Programs} in {Software} {Engineering}}.
\newblock
\newblock
\urldef\tempurl%
\url{https://www.acm.org/binaries/content/assets/education/se2014.pdf}
\showURL{%
\tempurl}


\bibitem[Joshi et~al\mbox{.}(2020)]%
        {joshi_evaluating_2020}
\bibfield{author}{\bibinfo{person}{Alark Joshi}, \bibinfo{person}{Marissa Schmidt}, \bibinfo{person}{Shane Panter}, {and} \bibinfo{person}{Amit Jain}.} \bibinfo{year}{2020}\natexlab{}.
\newblock \showarticletitle{Evaluating the {Benefits} of {Team}-{Based} {Learning} in a {Systems} {Programming} {Class}}. In \bibinfo{booktitle}{\emph{2020 {IEEE} {Frontiers} in {Education} {Conference} ({FIE})}}. \bibinfo{pages}{1--7}.
\newblock
\urldef\tempurl%
\url{https://doi.org/10.1109/FIE44824.2020.9274183}
\showDOI{\tempurl}
\newblock
\shownote{ISSN: 2377-634X}.


\bibitem[Krehbiel et~al\mbox{.}(2017)]%
        {krehbiel_agile_2017}
\bibfield{author}{\bibinfo{person}{Timothy~C Krehbiel}, \bibinfo{person}{Peter~A Salzarulo}, \bibinfo{person}{Michelle~L Cosmah}, \bibinfo{person}{John Forren}, \bibinfo{person}{Gerald Gannod}, \bibinfo{person}{Douglas Havelka}, \bibinfo{person}{Andrea~R Hulshult}, {and} \bibinfo{person}{Jeffrey Merhout}.} \bibinfo{year}{2017}\natexlab{}.
\newblock \showarticletitle{Agile {Manifesto} for {Teaching} and {Learning}}.
\newblock   \bibinfo{volume}{17} (\bibinfo{year}{2017}).
\newblock


\bibitem[Loftsson and Matthíasdóttir(2019)]%
        {loftsson_using_2019}
\bibfield{author}{\bibinfo{person}{Hrafn Loftsson} {and} \bibinfo{person}{Ásrún Matthíasdóttir}.} \bibinfo{year}{2019}\natexlab{}.
\newblock \showarticletitle{Using {Flipped} {Classroom} and {Team}-{Based} {Learning} in a {First}-{Semester} {Programming} {Course}: {An} {Experience} {Report}}. In \bibinfo{booktitle}{\emph{2019 {IEEE} {International} {Conference} on {Engineering}, {Technology} and {Education} ({TALE})}}. \bibinfo{pages}{1--6}.
\newblock
\urldef\tempurl%
\url{https://doi.org/10.1109/TALE48000.2019.9225985}
\showDOI{\tempurl}
\newblock
\shownote{ISSN: 2470-6698}.


\bibitem[Matalonga et~al\mbox{.}(2017)]%
        {matalonga_deploying_2017}
\bibfield{author}{\bibinfo{person}{Santiago Matalonga}, \bibinfo{person}{Gastón Mousqués}, {and} \bibinfo{person}{Alejandro Bia}.} \bibinfo{year}{2017}\natexlab{}.
\newblock \showarticletitle{Deploying {Team}-{Based} {Learning} at {Undergraduate} {Software} {Engineering} {Courses}}. In \bibinfo{booktitle}{\emph{2017 {IEEE}/{ACM} 1st {International} {Workshop} on {Software} {Engineering} {Curricula} for {Millennials} ({SECM})}}. \bibinfo{pages}{9--15}.
\newblock
\urldef\tempurl%
\url{https://doi.org/10.1109/SECM.2017.2}
\showDOI{\tempurl}


\bibitem[Michaelsen et~al\mbox{.}(2002)]%
        {michaelsen_team-based_2002}
\bibfield{editor}{\bibinfo{person}{Larry~K. Michaelsen}, \bibinfo{person}{Arletta~Bauman Knight}, {and} \bibinfo{person}{L.~Dee Fink}} (Eds.). \bibinfo{year}{2002}\natexlab{}.
\newblock \bibinfo{booktitle}{\emph{Team-{Based} {Learning}: {A} {Transformative} {Use} of {Small} {Groups}} (\bibinfo{edition}{1st edition} ed.)}.
\newblock \bibinfo{publisher}{Praeger}, \bibinfo{address}{Westport, Conn}.
\newblock
\showISBNx{978-0-89789-863-8}


\bibitem[Michaelsen and Sweet(2008)]%
        {michaelsen_essential_2008}
\bibfield{author}{\bibinfo{person}{Larry~K. Michaelsen} {and} \bibinfo{person}{Michael Sweet}.} \bibinfo{year}{2008}\natexlab{}.
\newblock \showarticletitle{The essential elements of team-based learning}.
\newblock \bibinfo{journal}{\emph{New Directions for Teaching and Learning}} \bibinfo{volume}{2008}, \bibinfo{number}{116} (\bibinfo{date}{Sept.} \bibinfo{year}{2008}), \bibinfo{pages}{7--27}.
\newblock
\showISSN{02710633, 15360768}
\urldef\tempurl%
\url{https://doi.org/10.1002/tl.330}
\showDOI{\tempurl}


\bibitem[Nascimento et~al\mbox{.}(2019)]%
        {nascimento_does_2019}
\bibfield{author}{\bibinfo{person}{Debora Maria~Coelho Nascimento}, \bibinfo{person}{Christina von Flach Garcia Chavez}, {and} \bibinfo{person}{Roberto~Almeida Bittencourt}.} \bibinfo{year}{2019}\natexlab{}.
\newblock \showarticletitle{Does {FLOSS} in {Software} {Engineering} {Education} {Narrow} the {Theory}-{Practice} {Gap}? {A} {Study} {Grounded} on {Students}’ {Perception}}. In \bibinfo{booktitle}{\emph{Open {Source} {Systems}}} \emph{(\bibinfo{series}{{IFIP} {Advances} in {Information} and {Communication} {Technology}})}, \bibfield{editor}{\bibinfo{person}{Francis Bordeleau}, \bibinfo{person}{Alberto Sillitti}, \bibinfo{person}{Paulo Meirelles}, {and} \bibinfo{person}{Valentina Lenarduzzi}} (Eds.). \bibinfo{publisher}{Springer International Publishing}, \bibinfo{address}{Cham}, \bibinfo{pages}{153--164}.
\newblock
\showISBNx{978-3-030-20883-7}
\urldef\tempurl%
\url{https://doi.org/10.1007/978-3-030-20883-7_14}
\showDOI{\tempurl}


\bibitem[Nawabi et~al\mbox{.}(2021)]%
        {nawabi_team-based_2021}
\bibfield{author}{\bibinfo{person}{Shazia Nawabi}, \bibinfo{person}{Rabia Bilal}, {and} \bibinfo{person}{Muhammad~Qasim Javed}.} \bibinfo{year}{2021}\natexlab{}.
\newblock \showarticletitle{Team-based learning versus {Traditional} lecture-based learning: {An} investigation of students’ perceptions and academic achievements}.
\newblock \bibinfo{journal}{\emph{Pakistan Journal of Medical Sciences}} \bibinfo{volume}{37}, \bibinfo{number}{4} (\bibinfo{date}{May} \bibinfo{year}{2021}).
\newblock
\showISSN{1681-715X}
\urldef\tempurl%
\url{https://doi.org/10.12669/pjms.37.4.4000}
\showDOI{\tempurl}
\newblock
\shownote{Number: 4}.


\bibitem[Ouhbi and Pombo(2020)]%
        {ouhbi_software_2020}
\bibfield{author}{\bibinfo{person}{Sofia Ouhbi} {and} \bibinfo{person}{Nuno Pombo}.} \bibinfo{year}{2020}\natexlab{}.
\newblock \showarticletitle{Software {Engineering} {Education}: {Challenges} and {Perspectives}}. In \bibinfo{booktitle}{\emph{2020 {IEEE} {Global} {Engineering} {Education} {Conference} ({EDUCON})}}. \bibinfo{pages}{202--209}.
\newblock
\urldef\tempurl%
\url{https://doi.org/10.1109/EDUCON45650.2020.9125353}
\showDOI{\tempurl}
\newblock
\shownote{ISSN: 2165-9567}.


\bibitem[Pinto et~al\mbox{.}(2019)]%
        {pinto_training_2019}
\bibfield{author}{\bibinfo{person}{Gustavo Pinto}, \bibinfo{person}{Clarice Ferreira}, \bibinfo{person}{Cleice Souza}, \bibinfo{person}{Igor Steinmacher}, {and} \bibinfo{person}{Paulo Meirelles}.} \bibinfo{year}{2019}\natexlab{}.
\newblock \showarticletitle{Training {Software} {Engineers} {Using} {Open}-{Source} {Software}: {The} {Students}' {Perspective}}. In \bibinfo{booktitle}{\emph{2019 {IEEE}/{ACM} 41st {International} {Conference} on {Software} {Engineering}: {Software} {Engineering} {Education} and {Training} ({ICSE}-{SEET})}}. \bibinfo{pages}{147--157}.
\newblock
\urldef\tempurl%
\url{https://doi.org/10.1109/ICSE-SEET.2019.00024}
\showDOI{\tempurl}


\bibitem[Pinto et~al\mbox{.}(2017)]%
        {pinto_training_2017}
\bibfield{author}{\bibinfo{person}{Gustavo Henrique~Lima Pinto}, \bibinfo{person}{Fernando~Figueira Filho}, \bibinfo{person}{Igor Steinmacher}, {and} \bibinfo{person}{Marco~Aurelio Gerosa}.} \bibinfo{year}{2017}\natexlab{}.
\newblock \showarticletitle{Training {Software} {Engineers} {Using} {Open}-{Source} {Software}: {The} {Professors}' {Perspective}}. In \bibinfo{booktitle}{\emph{2017 {IEEE} 30th {Conference} on {Software} {Engineering} {Education} and {Training} ({CSEE}\&{T})}}. \bibinfo{pages}{117--121}.
\newblock
\urldef\tempurl%
\url{https://doi.org/10.1109/CSEET.2017.27}
\showDOI{\tempurl}
\newblock
\shownote{ISSN: 2377-570X}.


\bibitem[Postner et~al\mbox{.}(2019)]%
        {postner_impact_2019}
\bibfield{author}{\bibinfo{person}{Lori Postner}, \bibinfo{person}{Darci Burdge}, \bibinfo{person}{Heidi J.~C. Ellis}, \bibinfo{person}{Stoney Jackson}, {and} \bibinfo{person}{Gregory~W. Hislop}.} \bibinfo{year}{2019}\natexlab{}.
\newblock \showarticletitle{Impact of {HFOSS} on {Education} on {Instructors}}. In \bibinfo{booktitle}{\emph{Proceedings of the 2019 {ACM} {Conference} on {Innovation} and {Technology} in {Computer} {Science} {Education}}} \emph{(\bibinfo{series}{{ITiCSE} '19})}. \bibinfo{publisher}{Association for Computing Machinery}, \bibinfo{address}{New York, NY, USA}, \bibinfo{pages}{285--291}.
\newblock
\showISBNx{978-1-4503-6895-7}
\urldef\tempurl%
\url{https://doi.org/10.1145/3304221.3319765}
\showDOI{\tempurl}


\bibitem[Presler-Marshall et~al\mbox{.}(2023)]%
        {presler-marshall_improving_2023}
\bibfield{author}{\bibinfo{person}{Kai Presler-Marshall}, \bibinfo{person}{Sarah Heckman}, {and} \bibinfo{person}{Kathryn~T. Stolee}.} \bibinfo{year}{2023}\natexlab{}.
\newblock \showarticletitle{Improving {Grading} {Outcomes} in {Software} {Engineering} {Projects} {Through} {Automated} {Contributions} {Summaries}}. In \bibinfo{booktitle}{\emph{2023 {IEEE}/{ACM} 45th {International} {Conference} on {Software} {Engineering}: {Software} {Engineering} {Education} and {Training} ({ICSE}-{SEET})}}. \bibinfo{pages}{259--270}.
\newblock
\urldef\tempurl%
\url{https://doi.org/10.1109/ICSE-SEET58685.2023.00030}
\showDOI{\tempurl}
\newblock
\shownote{ISSN: 2832-7578}.


\bibitem[Rands and Gansemer-Topf(2017)]%
        {rands_room_2017}
\bibfield{author}{\bibinfo{person}{Melissa Rands} {and} \bibinfo{person}{Ann Gansemer-Topf}.} \bibinfo{year}{2017}\natexlab{}.
\newblock \showarticletitle{The {Room} {Itself} {Is} {Active}: {How} {Classroom} {Design} {Impacts} {Student} {Engagement}}.
\newblock \bibinfo{journal}{\emph{Journal of Learning Spaces}} \bibinfo{volume}{6}, \bibinfo{number}{1} (\bibinfo{date}{Jan.} \bibinfo{year}{2017}), \bibinfo{pages}{26--33}.
\newblock
\urldef\tempurl%
\url{https://lib.dr.iastate.edu/edu_pubs/49}
\showURL{%
\tempurl}


\bibitem[Santos et~al\mbox{.}(2022)]%
        {santos_hits_2022}
\bibfield{author}{\bibinfo{person}{Italo Santos}, \bibinfo{person}{Igor Wiese}, \bibinfo{person}{Igor Steinmacher}, \bibinfo{person}{Anita Sarma}, {and} \bibinfo{person}{Marco~A. Gerosa}.} \bibinfo{year}{2022}\natexlab{}.
\newblock \showarticletitle{Hits and {Misses}: {Newcomers}' ability to identify {Skills} needed for {OSS} tasks}. In \bibinfo{booktitle}{\emph{2022 {IEEE} {International} {Conference} on {Software} {Analysis}, {Evolution} and {Reengineering} ({SANER})}}. \bibinfo{pages}{174--183}.
\newblock
\urldef\tempurl%
\url{https://doi.org/10.1109/SANER53432.2022.00032}
\showDOI{\tempurl}
\newblock
\shownote{ISSN: 1534-5351}.


\bibitem[Schön et~al\mbox{.}(2023)]%
        {schon_shift_2023}
\bibfield{author}{\bibinfo{person}{Eva-Maria Schön}, \bibinfo{person}{Ilona Buchem}, \bibinfo{person}{Stefano Sostak}, {and} \bibinfo{person}{Maria Rauschenberger}.} \bibinfo{year}{2023}\natexlab{}.
\newblock \showarticletitle{Shift {Toward} {Value}-{Based} {Learning}: {Applying} {Agile} {Approaches} in {Higher} {Education}}. In \bibinfo{booktitle}{\emph{Web {Information} {Systems} and {Technologies}}} \emph{(\bibinfo{series}{Lecture {Notes} in {Business} {Information} {Processing}})}, \bibfield{editor}{\bibinfo{person}{Massimo Marchiori}, \bibinfo{person}{Francisco~José Domínguez~Mayo}, {and} \bibinfo{person}{Joaquim Filipe}} (Eds.). \bibinfo{publisher}{Springer Nature Switzerland}, \bibinfo{address}{Cham}, \bibinfo{pages}{24--41}.
\newblock
\showISBNx{978-3-031-43088-6}
\urldef\tempurl%
\url{https://doi.org/10.1007/978-3-031-43088-6_2}
\showDOI{\tempurl}


\bibitem[Smith et~al\mbox{.}(2014)]%
        {smith_selecting_2014}
\bibfield{author}{\bibinfo{person}{Therese~Mary Smith}, \bibinfo{person}{Robert McCartney}, \bibinfo{person}{Swapna~S. Gokhale}, {and} \bibinfo{person}{Lisa~C. Kaczmarczyk}.} \bibinfo{year}{2014}\natexlab{}.
\newblock \showarticletitle{Selecting open source software projects to teach software engineering}. In \bibinfo{booktitle}{\emph{Proceedings of the 45th {ACM} technical symposium on {Computer} science education}} \emph{(\bibinfo{series}{{SIGCSE} '14})}. \bibinfo{publisher}{Association for Computing Machinery}, \bibinfo{address}{New York, NY, USA}, \bibinfo{pages}{397--402}.
\newblock
\showISBNx{978-1-4503-2605-6}
\urldef\tempurl%
\url{https://doi.org/10.1145/2538862.2538932}
\showDOI{\tempurl}


\bibitem[Spinellis(2021)]%
        {spinellis_why_2021}
\bibfield{author}{\bibinfo{person}{Diomidis Spinellis}.} \bibinfo{year}{2021}\natexlab{}.
\newblock \showarticletitle{Why computing students should contribute to open source software projects}.
\newblock \bibinfo{journal}{\emph{Commun. ACM}} \bibinfo{volume}{64}, \bibinfo{number}{7} (\bibinfo{date}{June} \bibinfo{year}{2021}), \bibinfo{pages}{36--38}.
\newblock
\showISSN{0001-0782}
\urldef\tempurl%
\url{https://doi.org/10.1145/3437254}
\showDOI{\tempurl}


\bibitem[van Helden et~al\mbox{.}(2023)]%
        {van_helden_collaborative_2023}
\bibfield{author}{\bibinfo{person}{Gitte van Helden}, \bibinfo{person}{Barry T.~C. Zandbergen}, \bibinfo{person}{Marcus~M. Specht}, {and} \bibinfo{person}{Eberhard K.~A. Gill}.} \bibinfo{year}{2023}\natexlab{}.
\newblock \showarticletitle{Collaborative {Learning} in {Engineering} {Design} {Education}: {A} {Systematic} {Literature} {Review}}.
\newblock \bibinfo{journal}{\emph{IEEE Transactions on Education}} (\bibinfo{year}{2023}), \bibinfo{pages}{1--0}.
\newblock
\showISSN{1557-9638}
\urldef\tempurl%
\url{https://doi.org/10.1109/TE.2023.3283609}
\showDOI{\tempurl}
\newblock
\shownote{Conference Name: IEEE Transactions on Education}.


\end{thebibliography}




\end{document}